\documentclass[12pt]{article}
\usepackage{epsfig}
\begin{document}

\title{Is the Universe Homogeneous on Large Scales?}

\author{Marc Davis\\
Depts. of Astronomy and Physics\\
University of California, Berkeley}

\maketitle

\section{Two Competing Visions}

\subsection{Introduction}

This morning's debate is focused on the question of the large-scale
homogeneity of the Universe.  I shall present the affirmative
position, that there is overwhelming evidence for large scale
homogeneity on scales in excess of approximately $50h^{-1}$ Mpc, with
a fractal distribution of matter on smaller scales.  My worthy
opponent, Dr.  Luciano Pietronero, will present the counter-argument,
that the fractal distribution observed on smaller scales continues to
the largest observed scales, and that there is no evidence for
homogeneity on any scale.

This is an important question, since the Friedmann-Robertson-Walker
metric presumes large scale homogeneity and isotropy for the Universe.
This is the simplest cosmological model, and it is a fair question to
ask the degree to which it is supported by the observational evidence.
We know that the galaxy distribution is far from homogeneous on small
scales, and  large-scale structure in the form of
long filaments and chains extends to lengths in excess of 100$h^{-1}$
Mpc \cite{dacosta94}.  Redshift surveys often show structures
nearly as large as the entire survey size; how confident can we be that 
homogeneity is a valid concept for the large-scale universe?

\subsection{The Mean Density of Galaxies}

The question of large scale homogeneity 
revolves around the issue of whether the mean density of
the Universe, $\bar n$, is a well defined concept. In the standard cosmological
model,  $\bar n$ is perfectly well defined, but in the
fractal model of the Universe, $\bar n$ has vanishing global value and 
has relevance only within a given survey volume.
In the conventional picture of large scale structure, 
the {\it rms} value of the number density of galaxies as observed at radius
$r$ from
the vantage point of any randomly chosen galaxy, $n(r)$, 
 is given from the usual
definition of the two-point galaxy correlation, $\xi(r)$, by
 $n(r)_{rms} = {\bar n}\left(1 + \xi(r)\right)$.
The measured $\xi(r)$ is well
characterized by a power law, $\xi(r) \approx (r/r_0)^{-\gamma}$,
($r_0 \approx 5h^{-1}$ Mpc, $\gamma = 1.8$).  Alternatively, this can
be expressed in terms of the power spectrum of density fluctuations,
$P(k) \propto k^{-1.2}$, $(\delta M/M)_{rms} \propto (k^3P(k))^{1/2}
=1$ at some well defined scale.  In the conventional picture, there
exists a distinct transition between ``small" and ``large" amplitude
scales, depending on whether $\xi(r) < 1$ or $\xi(r) > 1$.  On large
scales, where $\xi(r) \rightarrow 0$, one has simply $n(r) \equiv
M(r)/V(r) \rightarrow constant$.  On scales where the rms density
behaves as a power law, the galaxy distribution has fractal
properties, but on larger scales the galaxy distribution approaches
uniformity.

In contrast, within the fractal model advocated by Pietronero and
colleagues  \cite{piet87} 
\cite{baryshev} \cite{sl95}, there is {\it no} constant term in the
rms density, the mass within a radius $r$ scales as $M(r) \propto
r^D$, $D=3-\gamma \approx 1.2-1.5$, and $n(r)_{rms} = M(r)/V(r) \propto
r^{D-3} \rightarrow 0$ as $r\rightarrow \infty$.  
In order to preserve the
cosmological principle that we are not at the center of the Universe
or at any point of special symmetry, such an rms density behavior is
possible only if the galaxy distribution viewed from a typical
observer is extremely anisotropic, with 100\% density fluctuations at
any distance $r$ from a typical observer.  Such a universe does not
approach homogeneity on large scales, and large voids, a constant
fraction of the survey radius, would be expected within any survey.
Because one must always estimate the mean density from within the same
volume used to estimate the correlation function, in the fractal
picture one expects the correlation length to scale linearly as the
survey size $r$, with $r_0 \approx 0.4r$.  The average density has no
meaning in this universe; on a global scale the Universe is
empty and its average density
approaches zero.  There is no transition between linear and nonlinear
fluctutions:  $(\delta M / M)_{rms} \approx 1$ on all scales because
of the trend of mean density with sample volume.

Such a model is a radical departure from the standard model of a
homogeneous large-scale Universe.  Does the observational data support
such a conjecture, or can it be decisively ruled out?  I shall argue
that we have an abundance of observational
arguments in support of the standard
picture which are incompatible with the fractal model extended to
arbitarily large scale.  (The theoretical underpinning of such a radical
model is another matter altogether.)
The question of real interest is to determine
the outer scale of the fractal scaling behavior, and I shall argue
that there exist firm bounds on this scale.

\section{Evidence for Homogeneity (2-d tests)}

The largest data sets available are two-dimensional in nature, and
most of these have been well known for decades.  The remarkable
isotropy of the CMBR, X-ray counts, radio source counts, and the
$\gamma$-ray bursts all argue that we are either at the center of the
Universe or that on the largest scales the Universe is homogeneous.
The arguments given by Weinberg \cite{weinberg} are still valid today.
We know the CMBR comes to us from redshifts $Z \gg 1$, while the
discrete radio sources are distributed to $Z>1$, with the X-ray
background presumably arising from discrete sources at $Z < 3$.  If
the matter distribution is a pure fractal in space, how does it become
so smooth in projection on the sky?  The proponents of the fractal
universe argue that projection of the 3-d fractals is complicated and
that all structure can be lost.  It is well known that the projection
will dilute the information, but it seems evident that not {\it all}
the information would be lost and that isotropy at the remarkable
levels observed, e.g.  1\% precision, is not possible unless the outer
scale of the fractal structure is considerably smaller than the
redshift limit of the databases. Peebles \cite{pjep92} shows that
the large-scale isotropy of the X-ray background radiation constrains
the fractal dimension $D$ to be $|3-D| \le 0.001$ on large scales,
which
would seem rather definitive.  The proponents of the fractal
universe have been challenged to produce an example of a fractal that
projects to a uniform sky distribution, but to date they have failed
to do so.  

 Similarly, the observed counts of galaxies versus flux $f$, for
intermediate magnitudes in the range $14<m<18$, scales as $N(>f)
\propto f^{-3/2}$, just as expected in a homogeneous, Euclidean
Universe.  For fainter magnitudes, we observe $N(>f) \propto f^{-1}$,
which suggests a combination of evolutionary and expansion effects,
while the isotropy of the faint number counts is inconsistent with fractal
behavior.  Peebles \cite{pjep92}\cite{pjep80} presents considerable
detail on the constraints these arguments set
 on the fractal dimension $D$.   In
the fractal model, since $M(r) \propto r^{3-\gamma}$, we expect $N(>f)
\propto f^{(\gamma-3)/2} \propto f^{-0.6}$, which is very far from the
observations.  To fit the faint counts, one must adjust $\gamma$, but
this is inconsistent with the isotropy of the faint counts and with
the fractal dimension found on smaller scales.

The angular correlation function $w(\theta)$ of the large
two-dimensional catalogs of galaxies such as the APM catalog
\cite{Maddox} obeys the Limber scaling law to a remarkable degree
\cite{pjep92}.  Again this relationship is fully consistent with
homogeneity on large scale.  In a fractal universe, the angular
correlation length of a galaxy catalog should be a constant, large
fraction of the angular extent of the survey, and this scale should be
independent of the flux.  Again, this is exactly contrary to the
observations.  The proponents of the fractal model argue that the
projection of the three dimensional structure to the observed
$w(\theta)$ has erased all the information, but years of experience
with deconvolution demonstrate that recovery of $\xi(r)$ from
$w(\theta)$ is reliable and conceptually straightforward.  Recent
explicit constructions of fractal models \cite{pjep96} demonstrate
this point very clearly-- a three dimensional fractal leads to a very
anisotropic two-dimensional galaxy distribution.

\section{Evidence for Homogeneity (3-d tests)}

The past decade has witnessed the explosive growth of redshift surveys
of galaxies, from which one can estimate three-dimensional statistics
directly.  To date, these surveys are necessarily much smaller than
the very large databases such as the APM catalog, and they
correspondingly show more fluctuations from sample to sample.  It is
important that one beware that many of the early redshift surveys were
too small to represent a fair statistical sample of the Universe (as
expected in the orthodox school).  Furthermore, many of the existing
surveys are based on catalogs with irregular edges, and /or with known
non-uniformity in their sample selection.  These catalogs are usually
flux limited, and some of them are far from complete at any flux
level!  For example, the CfA2 survey is based on Zwicky magnitudes,
where systematic errors from one section of the sky to another are
suspected, the Perseus-Pisces redshift survey contains a region of
substantial extinction from our galaxy, the selection of Abell
clusters of galaxies has known selection effects that depend on the
zenith angle of the photographic plates from which the clusters are
selected.  But at least these catalogs are approximately complete.
Worst of all is to use a database such as LEDA \cite{paturel} or ZCAT 
which are repositories of redshifts collected from the literature.
There is no way to make a uniform correction for the completeness of
these catalogs, since they are constructed in an uncontrolled manner
and are most definitely {\it not} intended for statistical analysis.

The trouble this can lead to is exhibited in a power spectral analysis
of the LEDA sample \cite{amendola}.  A comparison of this measure of
$P(k)$ with that obtained from the complete CfA2 survey is shown in
the figures which Professor Pietronero prepared for this debate.  The
LEDA-derived $P(k)$ is an order of magnitude larger than that derived
from CfA2 and does not smoothly extrapolate to the Sachs-Wolfe
inferred power spectrum on the larger scales measured by COBE (while
the CfA2 spectrum can be smoothly connected to the COBE results).
Proponents of the fractal picture would cite this instability of
$P(k)$ as a demonstration of fractal scaling, but I would counter that
it is the LEDA database which is a fractal, not the Universe.

Redshift catalogs which are appropriate for statistical analysis are
those for which the selection is well known and well defined.  For the
purposes of today's debate, the most suitable redshift catalogs
presently available are the survey of a bright subset of the APM
galaxies \cite{loveday}, the IRAS catalogs \cite{fisher95}, and the
recently completed LCRS survey \cite{lin96}.

The notion that the galaxy distribution is a fractal arose from
analysis of the two point correlation function $\xi(r)$ in early
redshift surveys.  Many authors have commented that large coherent
structures often appear to be as large as they could be within the
sample volume surveyed (e.g.  the CfA stickman, \cite{dacosta94}).
There is widespread agreement that the galaxy distribution
approximates a fractal over a considerable range of scales
\cite{borgani}.  The early surveys displayed increasing correlation
length with increasing sample volume \cite{davis88}, and all surveys
continue to show stronger correlation amplitudes for rich clusters,
different correlation properties for different types of galaxies
\cite{Davis-Geller}, and weak evidence for increased correlation
strength for more luminous galaxies \cite{park}.  In the standard
model, these properties are explained by the ``bias" of rare events,
luminosity bias, and environmental effects.

The defense of the standard interpretation of a homogeneous
large-scale universe has never been based on examination of the
stability of $\xi(r)$ or on examination of redshift survey maps; its
defense up until the recent data has rested on the isotropy of the 
two-dimensional catalogs and on the stability of the
mean density as a function of redshift and direction, $n(z,\omega)$.
Local surveys in opposite hemispheres have very similar $n(z)$ curves,
and even in the original CfA1 survey, the mean galaxy 
density derived in the
Northern galactic hemisphere agreed with that in the Southern
hemisphere to within a factor of 1.3  \cite{DH82}.

In the fractal interpreatation, this ``bias" explanation for the
behavior of $\xi(r)$ is considered 
a cop-out.  Instead, one argues that $n(r)$ is poorly
defined and the increased $\xi(r)$ for larger volumes is simply the
result of a decreasing mean density,  as expected in the
fractal model.  The problems with this explanation are two-fold.  In the
fractal model the correlation length 
$r_0$ should increase linearly with sample depth, but
even in the shallowest slices of the 
 CfA1 survey $r_0$ increases only as the
square-root of the sample depth \cite{davis88}.  Furthermore, it has
been clearly demonstrated that different populations of galaxies drawn
from the {\it same} volume do exhibit differing correlation properties
\cite{park}, so therefore biasing effects must exist.

\section{Recent Results (using 3-d data)}

\subsection{IRAS redshift surveys}

Pietronero for years has argued that the most suitable volume for
statistical analysis is a sphere, since it most efficiently contains
the largest fraction of galaxies within the most compact surface.
This ideal has now been closely approximated by the IRAS selected
galaxy samples, \cite{fisher95} which cover 88\% of the sky to a depth
of roughly 180$h^{-1}$ Mpc, although the diluteness is quite extreme
beyond 80 $h^{-1}$ Mpc.  Full sky maps of the observed galaxy
distribution of the 1.2 Jy IRAS survey are shown in Figure 1.  Each of
these plots are independent slices of redshift with nearly
constant aspect ratio, such that $\Delta z/z
\approx 1$.  

In an unbounded fractal universe, the galaxy distribution must be extremely
anisotropic if we are not at a special location.  The most elementary 
aspect of a fractal is that it should be approximately scale invariant,
which implies that
 all four of these figures of nearly constant aspect ratio should
appear statistically very similar to each other.   But the reality is
very different.
 
Within the IRAS
survey, or any optical survey, the galaxy distribution
 in the nearest shell, $cz<
1600$ km/s, is characterised by 100\% fluctuations from one hemisphere to
the other.  This is the expected behavior of a fractal distribution. 
But the more distant shells are progressively more and more
isotropic.  Here is a clear demonstration of the fractal behavior on
small scales giving way to homogeneity on large scales, and it is
completely contrary to the scale-invariant fractal picture, in which
{\it each} shell should exhibit similar 100\% anisotropies.  Plots of $n(z)$
for one hemisphere compared to one another show complete consistency
on large scales, indicating that there is no ambiguity in the
definition of the mean density.  Analysis of $\xi(r)$ for four
separate volume limited subsets of the 1.2 Jy IRAS survey yields a
correlation length $r_0 \approx 4h^{-1}$ Mpc that does not change as
the volume limiting radius is increased from 60$h^{-1}$ Mpc up to 120
$h^{-1}$ Mpc \cite{fisher}.  The disagreement with the fractal
predicted amplitude $r_0$ ranges from a factor of 6 to 12.

Professor Pietronero, in this conference, stated that he agrees that
the IRAS survey does not exhibit fractal behavior, and he ascribes
this to the diluteness of the sample.  But the problem with the
fractal model is that the maps shown in Figure 1 are too smooth;
dilute sampling could have increased the fluctuations, but how could
it have transformed an anisotropic map into a smooth, isotropic map?
The maps show that the outer scale of the fractal structure must be at
some radius within the second shell, which is much less anisotropic
than the first shell.  Thus, the conjecture that the Universe is a
fractal to the largest observed
scales is false, and the diluteness of the IRAS
sample cannot change this conclusion.

\subsection{Las Campanas Redshift Survey (LCRS)}

The LCRS \cite{lin96} is the first of the next generation of large,
deep redshift surveys.  It contains over 25,000 galaxies selected in
six strips of 80 by 1.5 degrees, to a depth $cz \approx 45,000$ km/s.
Three of these strips are nearly adjacent in the North, and three are
nearly adjacent in the South.

Recently published plots of the LCRS galaxy distribution show an ``end
of greatness".  There are many structures and voids as large as seen
in the shallower surveys such as CfA2, but there is an absence of
larger scale structure.  It appears as though the survey has crossed
the peak in the power spectrum of fluctuations $P(k)$ .  The six
slices of the LCRS all seem statistically very similar to each, and
the observed distributions of $n(z)$ in the different directions are
all the same on large scale, as expected in the standard picture but
quite contrary to the idea that the fractal scale extends to the full
survey depth.  If the fractal did extend to the full LCRS depth, the
separate slices should be very different from each other.  Thus LCRS,
consistent with IRAS, strongly demonstrates the approach to large
scale homogeneity.

\subsection{QSO Absorption Lines}

The spectra of all quasars at redshifts $z > 1.8$ contain abundant
Ly-$\alpha$ absorption lines due to intervening clouds of neutral
hydrogen.  In a fractal universe with no outer scale, most of the
quasars would have large voids in front of them, and so exhibit no
Ly-$\alpha$ absorption clouds.  A cursory glance at high quality
spectra of QSO's (e.g.  as taken by HIRES on Keck \cite{sargent})
shows that these clouds are ubiquitous, and that the universe is
statistically uniform in all directions probed.  The interval probed
by the clouds is roughly $1.8 < z < 4$, a comoving scale of
approximately $0.2c/H_0$.

There are no large holes in the distribution of the Ly-$\alpha$
clouds, and in fact they are so uniformly distributed that it is very
difficult to to measure any spatial correlations in the clouds at all
\cite{Webb}.  The Ly-$\alpha$ clouds appear to be very nearly uniformly
distributed in space, and are the most homogeneous tracers yet discovered.

All lines of sight are observed to be statistically equivalent, as
expected in a universe homogeneous on large scale.  Again, this is
completely contrary to the expectations of an unbounded fractal model.
It is quite clear that the outer scale of the fractal distribution of
matter must be orders of magnitude less than the 600$h^{-1}$ Mpc
probed by the sight line to the distant QSO's.

\section{Summary}

As I have briefly reviewed above, there exist numerous arguments which 
demonstrate that 
the outer scale of the fractal distribution is well within the scale
of observed  volumes.  From recent redshift space maps, we detect a
characteristic size of voids in the galaxy distribution consistent
with a peak in the power spectrum of fluctuations.  As emphasized by
Kirshner, the new LCRS survey may be seeing the ``end of greatness" of
large-scale structure.  Future surveys such as the AAT redshift survey
and the Sloan digital sky survey will lead to much better constraints
on the turnover of the large-scale power spectrum and on the
amplitude of the large-scale fluctuations.  Since I believe that our
current surveys are appoaching fair sample volumes, I fully expect the
correlation amplitudes of the future, massive surveys to be consistent
with current measurements.

The measured two-point galaxy correlation function $\xi(r)$ is a power
law over three decades of scale and approximates fractal behavior from
scales of $0.01h^{-1} < r < 10h^{-1}$ Mpc, but on scales larger than
$\approx 20h^{-1}$ Mpc, the fractal structure terminates, the {\it
rms} fluctuation amplitude falls below unity, and the Universe
approaches homogeneity, as necessary to make sense of a FRW universe.
There clearly exists an outer scale of the fractal behavior, and this
outer scale must grow with time (unless $\Omega_0 \ll 1$).  The observed
galaxy distribution, being a real physical system rather than a
mathematical idealization, is a beautiful example of a limited-scale
fractal joined onto sensible, dynamically evolving outer boundary
conditions.

\bigskip

As testimony to our faith in our respective debating positions, 
Professor Pietronero and I have agreed to 
wager a case of the best wine from Italy against the best Californian wine 
on the following
proposition:  that the correlation length $r_0$ as ultimately measured
from optically selected galaxies in
the Sloan digital sky 
survey will be larger (according to LP) or smaller (according
to MD) than $r_{0-CfA2}
\left(L_{Sloan}/L_{CfA2}\right)^{1/2}$, where $L_{survey}$ is the
radius of the largest inscribed sphere in a given survey.  (This
splits the harmonic difference of the fractal versus standard
prediction.)  Neil Turok has agreed to arbitrate this wager. Side bets
are welcome.









\vfill\eject

\begin{figure}[t]
\vskip -.5in
\epsfig{file=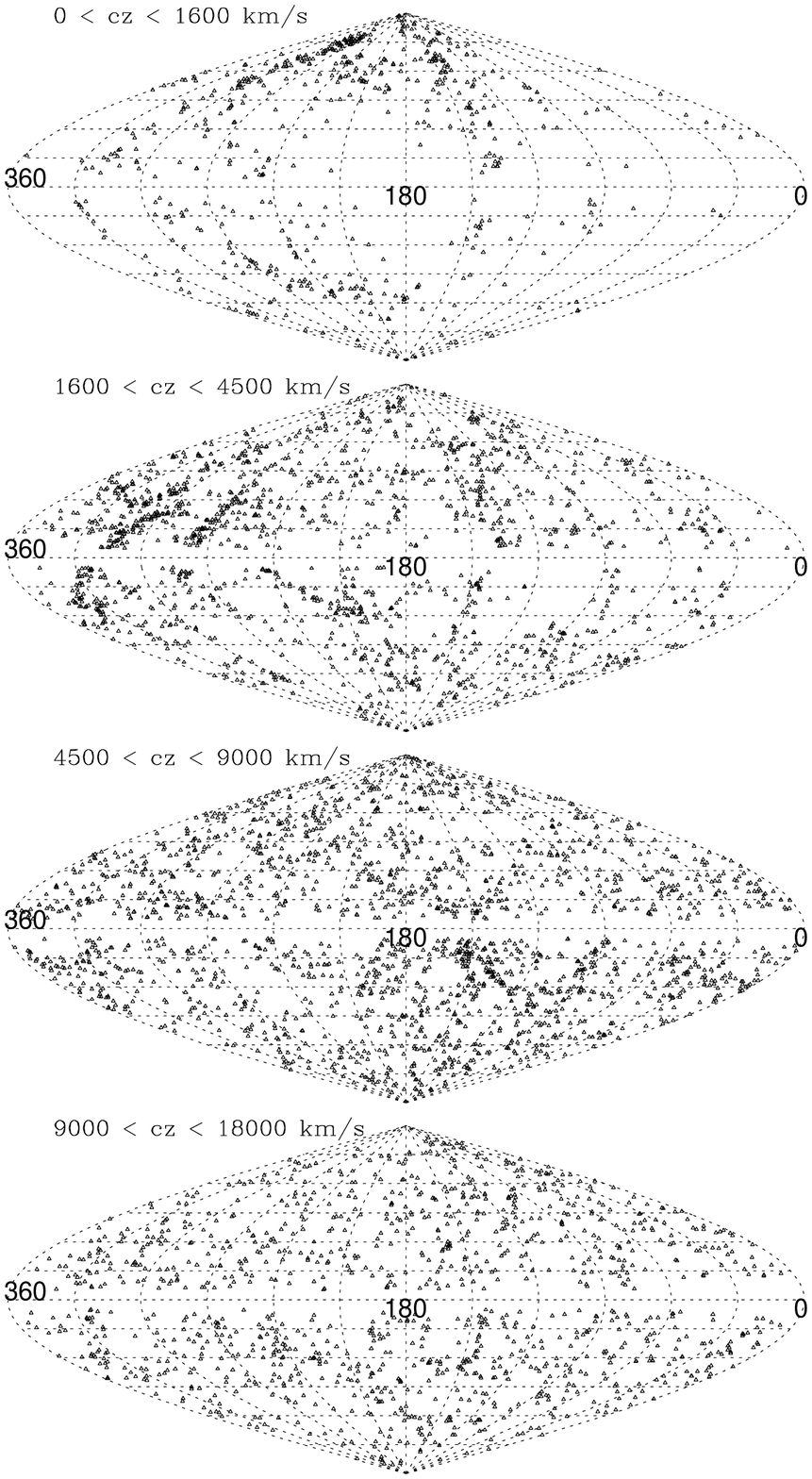, height=7in, angle=-0}
\caption{  Whole sky projection in galactic coordinates of the IRAS 1.2 Jy
redshift survey.  The different plots represent the galaxy distribution as
directly observed in different windows of observed redshift in the Local Group
frame.  Note the strong anisotropy in the nearby shell, progressively
diminishing in the distant shells.  }
\label{fig:1}
\end{figure}

\end{document}